\documentstyle[epsfig,psfig,ncs,12pt]{article}
\newcommand{\beq} { \begin{eqnarray} }
\def\bea{\begin{eqnarray}}
\def\eea{\end{eqnarray}}
\def\beq{\begin{equation}}
\def\eeq{\end{equation}}
\def\bef{\begin{figure}}
\def\eef{\end{figure}}
\def\bet{\begin{table}}
\def\eet{\end{table}}%
\def\eq#1{{eq.~(\ref{#1})}}

\def\fig#1{{Fig.~\ref{#1}}}
\def\fige#1#2{{Fig.~\ref{#1}{#2 }}}
\def\ra{\rightarrow}
\def\ot{\otimes}

\def\lsim{\raise0.3ex\hbox{$\;<$\kern-0.75em\raise-1.1ex\hbox{$\sim\;$}}}
\def\gsim{\raise0.3ex\hbox{$\;>$\kern-0.75em\raise-1.1ex\hbox{$\sim\;$}}}
\def\VEV#1{\left\langle #1\right\rangle}
\def\21{$SU(2) \ot U(1)$}
\def\321{$SU(3) \ot SU(2) \ot U(1)$}

\newcommand{\chip} [1] {\tilde{\chi}^{+}_{#1} }
\newcommand{\chipm} [1] {\tilde{\chi}^{\pm}_{#1} }
\newcommand{\chiz} [1] {\tilde{\chi}^{0}_{#1} }

\newcommand{\gev} { \hspace*{1mm} {\rm GeV} }
\newcommand{\mchip} [1] {m_{\chip{#1}} }
\newcommand{\mchiz} [1] {m_{\chiz{#1}} }
\newcommand{\mst} {m_{\tilde{t}_1} }
\newcommand{\no} { \nonumber \\ }
\newcommand {\stbt} {\tilde{t}_1 \to b + \tau }
\newcommand {\stch} [1] {\tilde{t}_1 \to c + {\tilde{\chi}^0_{#1}} }
\newcommand{\tanb} {\tan \beta}

\def\np#1#2#3{           {\it Nucl. Phys. }{\bf #1} (19#2) #3}
\def\pl#1#2#3{           {\it Phys. Lett. }{\bf #1} (19#2) #3}
\def\pr#1#2#3{           {\it Phys. Rev. }{\bf #1} (19#2) #3}
\def\prep#1#2#3{         {\it Phys. Rep. }{\bf #1} (19#2) #3}
\def\prl#1#2#3{          {\it Phys. Rev. Lett. }{\bf #1} (19#2) #3}

\def\n.c.#1#2#3{         {\it Nuovo Cim. }{\bf #1} (19#2) #3}
\def\r.n.c.#1#2#3{       {\it Riv. del Nuovo Cim. }{\bf #1} (19#2) #3}

\relax
\begin{document}
\thispagestyle{empty}
\begin{titlepage}
\begin{center}
\hfill FTUV/96-29\\
\hfill IFIC/96-33\\
\hfill HEPHY-PUB 647/96\\
\hfill UWThPh-1996-36\\
\vskip 0.4cm
{\Large \bf New signatures for a light stop at LEP2 in SUSY models
       with spontaneously broken R-parity}
\vskip 0.2cm
{\large A. Bartl}\footnote{E-mail bartl@pap.univie.ac.at}, 
{\large W. Porod} \footnote{E-mail porod@pap.univie.ac.at} \\
{\it Institut f\"ur Theoretische Physik, University of Vienna \\
A-1090 Vienna, Austria}\\
\vskip 0.2cm
{\large M. A. Garc\'{\i}a-Jare\~no}
\footnote{E-mail miguel@flamenco.ific.uv.es}, 
{\large M. B. Magro}\footnote{E-mail magro@flamenco.ific.uv.es} and
{\large J. W. F. Valle}
\footnote{E-mail valle@flamenco.ific.uv.es}\\
{\it Instituto de F\'{\i}sica Corpuscular - C.S.I.C.\\
Departament de F\'{\i}sica Te\`orica, Universitat de Val\`encia\\
46100 Burjassot, Val\`encia, SPAIN\\
 http://neutrinos.uv.es}\\
\vskip .2cm
{\large W. Majerotto} \footnote{E-mail majer@qhepu1.oeaw.ac.at} \\
{\it Institut f\"ur Hochenergiephysik, Akademie der Wissenschaften \\
A-1050 Vienna, Austria}\\
\vskip 0.2cm
{\bf Abstract}
\end{center}
\begin{quotation}
\noindent

In a class of supersymmetric models with R-parity breaking
the lightest stop can have new decay modes into third generation 
fermions, $\stbt$. We show that this decay may be 
dominant or at least comparable to the ordinary R-parity conserving 
mode  $\stch{1}$, where $\chiz{1}$ denotes the
lightest neutralino.  The new  R-parity violating decay modes could 
provide new signatures for stop production at LEP.

\end{quotation}
\end{titlepage}

\section{Introduction}

So far most searches for supersymmetric particles have so far
been performed within the framework of the {\sl Minimal Supersymmetric 
Standard Model (MSSM)} \cite{mssm}. Although the MSSM is by far the most
well studied realization of supersymmetry, there is considerable 
theoretical and phenomenological interest in studying possible 
implications of alternative scenarios \cite{beyond}, 
in which R-parity is broken. 
Indeed, neither gauge invariance nor supersymmetry
requires the conservation of R-parity. The violation of 
R-parity could arise  explicitly \cite{expl} as a residual 
effect of some larger unified theory \cite{expl0},
or spontaneously, through nonzero vacuum expectation values (VEV's)
  for scalar neutrinos \cite{aul}. In realistic  spontaneous R-parity
 breaking models \cite{MASIpot3,MASI,ROMA,ZR} 
the scale of R-parity violation lies
around the TeV scale. Its effects can therefore be 
large enough to be experimentally observable. Moreover,
these models are fully consistent with astrophysics and
cosmology \cite{beyond}.
In either case the supersymmetric (SUSY) particles need not be 
produced only in pairs, and the lightest of them could decay.

There are two generic cases of spontaneous R-parity
 breaking models to consider.
If lepton number is part of the gauge symmetry there is an
additional gauge boson which gets mass via the Higgs mechanism, and
there is no physical Goldstone boson \cite{ZR}. In this model
the lightest SUSY particle (LSP) is in general a neutralino which
decays mostly into visible states, therefore
breaking R-parity. The main decay modes are three-body decays such as
\beq
\label{vis}
\chiz{1} \ra f \bar{f} \nu,
\eeq
where f denotes a charged fermion. Its invisible decay modes
are in the channel
\beq
\label{invi}
\chiz{1} \ra 3 \nu.
\eeq
Alternatively, if spontaneous R-parity violation occurs
in the absence of any additional gauge symmetry, it leads to the
existence of a physical massless Nambu-Goldstone boson, called
majoron (J). Thus in this case {\sl the lightest SUSY
particle is the majoron} which is massless and therefore stable
\footnote{The majoron may have a small mass due to explicit
breaking effects at the Planck scale. In this case it may decay
into neutrinos and photons. However, the time scales are only of
cosmological interest and do not change the signal expected in
laboratory experiments \cite{KEV}.}. As a consequence the lightest
neutralino $\chiz{1}$ may decay invisibly as
\beq
\label{invis}
\chiz{1} \ra \nu + J,
\eeq
where the majoron is mainly a singlet \cite{MASIpot3,MASI}.
This last decay conserves R-parity since the majoron has a large
R-odd singlet sneutrino component.

In this paper we focus on the decay modes of the lightest top
squark in models where supersymmetry is realized with spontaneous 
R-parity violation. In such models the lightest stop could even be 
the lightest supersymmetric particle and be produced at LEP. 
Indeed present $e^+ e^-$ collider data \cite{LEPSEARCH} as well as 
$p\bar{p}$ collider data \cite{D0} do not preclude this possibility.

In supersymmetric models with spontaneous 
breaking of R-parity or in models, where this violation is parametrized 
explicitly through a bilinear superpotential term of the type $\ell H_u$ 
\cite{epsi}, the stop can have new decay modes such as 
\beq
\label{btau}
\stbt
\eeq
due to mixing between charged leptons and charginos.
In this paper we show that this decay may be dominant or at least 
comparable to the ordinary
 R-parity conserving mode
\beq
\label{chic}
\stch{1},
\eeq
where $\chiz{1}$ denotes the lightest neutralino.

\section{Lepton-Gaugino-Higgsino Mixing}

The basic tools in our subsequent discussion are the chargino
and neutralino mass matrices. The chargino mass matrix may be written as
\cite{MASIpot3}
\beq
\begin{array}{c|cccccccc}
& e^+_j & \tilde{H^+_u} & -i \tilde{W^+}\\
\hline
e_i & h_{e ij} v_d & - h_{\nu ij} v_{Rj} & \sqrt{2} g_2 v_{Li} \\
\tilde{H^-_d} & - h_{e ij} v_{Li} & \mu & \sqrt{2} g_2 v_d\\
-i \tilde{W^-} & 0 & \sqrt{2} g_2 v_u & M_2
\end{array}
\label{chino}
\eeq
Its diagonalization requires two unitary matrices U and V 
\bea
{\chi}_i^+ = V_{ij} {\psi}_j^+\\
{\chi}_i^- = U_{ij} {\psi}_j^- ,
\label{INO}
\eea
where the indices $i$ and $j$ run from $1$ to $5$ and
$\psi_j^+ = (e_1^+, e_2^+ , e_3^+ ,\tilde{H^+_u}, -i \tilde{W^+}$)
and $\psi_j^- = (e_1^-, e_2^- , e_3^-, \tilde{H^-_d}, -i \tilde{W^-}$).

While the details of the neutralino mass matrix are
rather model dependent, for our purposes it will be sufficient
to use the following effective form given by \cite{MASIpot3}
\beq
\begin{array}{c|cccccccc}
& {\nu}_i & \tilde{H}_u & \tilde{H}_d & -i \tilde{W}_3 & -i \tilde{B}\\
\hline
{\nu}_i & 0 & h_{\nu ij} v_{Rj} & 0 & g_2 v_{Li} & -g_1 v_{Li}\\
\tilde{H}_u & h_{\nu ij} v_{Rj} & 0 & - \mu & -g_2 v_u & g_1 v_u\\
\tilde{H}_d & 0 & - \mu & 0 & g_2 v_d & -g_1 v_d\\
-i \tilde{W}_3 & g_2 v_{Li} & -g_2 v_u & g_2 v_d & M_2 & 0\\
-i \tilde{B} & -g_1 v_{Li} & g_1 v_u & -g_1 v_d & 0 & M_1
\end{array}
\label{nino}
\eeq
This matrix is diagonalised by a $7 \times 7$ unitary matrix N,
\beq
{\chi}_i^0 = N_{ij} {\psi}_j^0 ,
\eeq
where
$\psi_j^0 = ({\nu}_i,\tilde{H}_u,\tilde{H}_d,-i \tilde{W}_3,-i \tilde{B}$),
with $\nu_i$ denoting the three weak-eigenstate neutrinos.

In the above  equations $v_u$ and $v_d$ are the VEV's responsible
for the breaking of the electroweak symmetry and the generation of
fermion masses, where the combination $v^2 = v_u^2 + v_d^2$
is fixed by the $W,Z$ masses. $v_R$ is the VEV mostly
responsible for the spontaneous violation of R-parity
\footnote{There is also a small seed of R-parity
breaking in the doublet sector, $v_L = \VEV {\tilde{\nu}_{L\tau}}$,
whose magnitude is related to the Yukawa coupling
$h_{\nu}$. Since this vanishes as $h_{\nu} \ra 0$,
we can naturally obey the limits from stellar energy
loss \cite{KIM}}. Moreover, $M_{1,2}$ denote the supersymmetry
breaking gaugino mass parameters and $g_{1,2}$ are the
\21 gauge couplings divided by $\sqrt{2}$, and we assume the
canonical relation $M_1/M_2 = \frac{5}{3} tan^2{\theta_W}$.
Note that the effective Higgsino mixing parameter $\mu$ may be
given in some models as $\mu = h_0 \VEV \Phi$, where $\VEV \Phi$
is the VEV of an appropriate singlet scalar.

Most of our subsequent analysis will be general enough to
cover a wide class of \21 models with spontaneously broken R-parity,
such as those of ref. \cite{MASIpot3,MASI}, as well as models
where the majoron is absent due to an enlarged gauge structure
\cite{ZR}. Many of the phenomenological features relevant
for the LEP studies discussed here already emerge in an effective
model, where the violation of R-parity is introduced explicitly
through a bilinear superpotential term of the type $\ell H_u$
\cite{epsi}. In this case the combination  $h_{\nu} v_R$ is
replaced by a mass parameter $\epsilon$.

Typical values of interest for LEP2 for the SUSY parameters $\mu$, $M_2$,
and the parameters $h_{\nu i,3}$ lie in the range given by
\beq
\label{param2}
\begin{array}{cccc}
-1000 \gev \leq \mu \leq 1000  \gev & & &
40 \gev \leq M_2 \leq 200 \gev \\
&&& \\
10^{-10}\leq h_{\nu 13},h_{\nu 23} \leq 10^{-1} & & &
10^{-5}\leq h_{\nu 33} \leq 10^{-1}\\
\end{array}
\eeq
while the expectation values can be chosen as
\beq
\label{param1}
\begin{array}{cccc}
v_L \equiv v_{L3}=100\:\: \mbox{MeV} & & &
v_{L1}=v_{L2}=0 \\
&&& \\
50\:\: \mbox{GeV}\leq v_R \equiv v_{R3}\leq 1000 \:\: \mbox{GeV} \ & & &
v_{R1}=v_{R2}=0\\
\end{array}
\eeq
with $1 \lsim \tan\beta=\frac{v_u}{v_d} \lsim 40$.

There are restrictions on these parameters that follow 
from searches for SUSY particles at LEP \cite{LEPSEARCH} and 
the TEVATRON \cite{D0}. In addition, we take into
account the constraints from  neutrino physics and 
weak interactions phenomenology \cite{bounds},
which are more characteristic of R-parity breaking models.
These are important, as they exclude 
many parameter choices that are otherwise allowed by the 
constraints from the collider data, while the converse is 
not true. Due to these constraints R-parity violation effects 
manifest themselves mainly in the third generation as expressed 
in \eq{param1}. In order to evaluate the masses and couplings of 
charginos and neutralinos we have performed a sampling of parameters 
in our model which are allowed by all constraints above. 

One finds that the  R-parity conserving
couplings for the lightest neutralino $\chiz{1}$ and the lightest 
chargino $\chipm{1}$ are of the same order as those in the MSSM. 
The R-parity violating chargino couplings may reach a few percent 
or so for chargino masses accessible at LEP \cite{bounds,bounds1}.
These couplings will 
induce a mixing between the charginos and the tau lepton, leading to
the R-parity violating decay in \eq{btau}. The corresponding decay 
width will be calculated below.

\section{ Decays of the Top Squark}

As we saw in the previous section, R-parity breaking effects should
occur mainly in the third generation, due to the mixing 
of the charged leptons with the charginos. As a result  the 
 decay mode $\tilde{t}_1 \to b + \tau$ is possible.
In the following we assume $\mst < m_b + \mchip{1}$.
In this case the most important R-parity conserving stop 
decay at LEP2 is $\tilde{t}_1 \to c + \tilde{\chi}^0_1$
\cite{lep2}.
Now the question arises whether the branching ratios of the decays into 
$b \tau$ and $c \tilde{\chi}^0_1$ are of comparable size.
As we will show $\stbt$ can be equally or even more important than
$\stch{1}$. We have also taken into account the decay $\stch{2}$, 
which is also possible in a small region of parameter space.
In the alternative case $\mst > m_b + \mchip{1}$
the decay mode $\tilde{t}_1 \to b + \chip{1}$ would clearly dominate the
decays (\ref{btau}) and (\ref{chic}). A detailed discussion of all R-parity
conserving stop decays at LEP2 is given in \cite{ourLEP}.

The decay $\stch{1}$ is only possible in higher order (see \fig{stopdecay}). 
As shown in \cite{HikKob} the dominant contribution is due to 
$\tilde{t}_1 \to \tilde{c}_L \to c \chiz{1}$. 
This can be parametrized by the following squark mass squared matrix 
in the ($\tilde{t}_L,\tilde{t}_R,\tilde{c}_L$) basis:
\beq
M^2 = \left( \begin{array}{ccc}
                        M^2_{\tilde{t}_L} & a_t m_t & \Delta_L \\
                        (a_t m_t)^* & M^2_{\tilde{t}_R} & \Delta_R \\
                        \Delta^*_L & \Delta^*_R & M^2_{\tilde{c}_L}
                       \end{array}
                 \right).
\eeq
Explicit expressions for $M^2_{\tilde{t}_L},M^2_{\tilde{t}_R},
a_t m_t,M^2_{\tilde{c}_L},\Delta_L$ and $\Delta_R$ can be found in
ref. \cite{HikKob}. Quite generally, flavour changing mass terms
are induced in supergravity models assuming unification and 
evolution according to the renormalization group equations \cite{Wyler}. 
The light stop eigenstate can therefore be written as:
\beq
\tilde{t}_1 = c_1 \tilde{t}_L + c_2 \tilde{t}_R + \delta \tilde{c}_L. 
\eeq
Estimates for $\delta$ were given in \cite{HikKob}. These are, 
however, very model dependent as pointed out in \cite{HalGab}.
Nevertheless a value $\delta \sim \sqrt{\frac{\Delta_{L,R}}{M_{\tilde{q}}^2}}$
may be regarded as typical. For our present phenomenological discussion 
we take a very strong $\tilde{t}$-$\tilde{c}$ mixing, $\delta \sim O(0.1)$. 
To be specific, we shall take $\delta = 0.1$.
Note that this is a conservative assumption, i.e. should  $\delta$ be smaller,
the relative importance of our novel decay \eq{btau} would be further enhanced.

The decay widths for (\ref{btau}) and (\ref{chic}) are: 
\bea
\Gamma(\stbt) &=&
 \frac{g^2}{16 \pi \mst^3} 
  \sqrt{(\mst^2-m_b^2-m_{\tau}^2)^2 - 4 m_b^2 m_{\tau}^2}
  \no
 & & \hspace{5mm}
    \times ( (l^2+k^2) (\mst^2-m_b^2-m_{\tau}^2) - 4 l k m_b m_{\tau} ) \\
\Gamma(\stch{i}) &=&
 \frac{g^2}{16 \pi \mst^3} |\delta|^2 f^2_i (\mst^2 - m_{\chiz{i}}^2)^2
\label{gamch}
\eea
where
\bea
l &=& \frac{m_t}{\sqrt{2} m_W \sin \beta} V_{34}  c_2 K_{tb}
   - V_{35} (c_1 K_{tb} + \delta K_{cb}), \hspace{5mm}
k = \frac{m_b K_{tb}}{\sqrt{2} m_W \cos \beta} U_{34} c_1
\label{coupk} \\
f_1 &=& - \frac{\sqrt{2}}{6} ( \tan \theta_W N_{47} + 3 N_{46} ), \hspace{5mm}
f_2 = - \frac{\sqrt{2}}{6} ( \tan \theta_W N_{57} + 3 N_{56} )
\label{coupf}
\eea
Here $V_{34}$, $U_{34}$ ($V_{35}$) denote the amount of tau-Higgsino
(tau-gaugino) mixing (see \eq{chino} and \eq{INO}) and 
$K_{tb} = 0.999, K_{cb} = 0.04$ are the corresponding CKM matrix 
elements \cite{PartDat}.

For a dominantly Higgsino-like neutralino \cite{Bartl89} (small 
$N_{i6}$ and $N_{i7}$) the R-parity conserving decay is induced
by the charm Yukawa coupling and is therefore suppressed
(see \eq{gamch} and \eq{coupf}). As a result,
the branching ratio of the R-parity 
violating $b \tau$ mode will be enhanced. 

In our present numerical analysis we have fixed 
$\mst = 80$ GeV, $c_1 = 0.5$, $v_R = 100$ GeV, $v_L = 0.1$ GeV and 
$h_{\nu 33} = 0.03$. We have varied the other parameters in the range
given in \eq{param1} and \eq{param2}. As already mentioned, we have
taken into account all bounds from the LEP experiments as well as 
neutrino physics \cite{bounds}
\footnote{Note that the MSSM LEP limits on neutralino masses do not hold
in broken R-parity models. For a recent discussion see ref. \cite{pimenta}.}.
In Fig. 2  we show the maximum value of the branching ratio for 
$\stbt$ in per cent as a function of the lightest neutralino mass 
$\mchiz{1}$ in the range $1\leq \tan\beta \leq 40$.   
One can see that BR($\stbt$) can easily 
reach 80~\% for neutralino masses of 50~GeV. 
There are a few points, not shown in Fig.~2, which lie 
{\sl above} the "maximum curve" even for 
relatively small neutralino masses. They correspond to "fine-tuned"
situations where $f_1$ almost vanishes due to a cancellation
between the terms in \eq{coupf} for small $\tan\beta$ values. 

Particularly interesting is the dependence
on $\tanb$, which can be seen in \fige{figtgbeta}.
The most characteristic feature is the existence of
a strong correlation between the branching ratio for
the $b\tau$ stop decay mode and  $\tan\beta$,
apart from a small region
in parameter space characterized by small $\tan\beta$ 
where $f_1$ vanishes and the R-parity conserving decay is 
suppressed
\footnote{These points correspond precisely to those we
have eliminated from Fig.~2 as mentioned above.}. 
 One can clearly see that most points with a large
branching ratio of $\stbt$ have a large $\tanb$.
Note that in the framework of the MSSM with 
radiative electroweak breaking such large $\tanb$
models have been advocated \cite{pokorski}.
The trend towards large $\tanb$ 
can be understood as being due to the influence of the 
large bottom Yukawa coupling (see \eq{coupk}). It is
also interesting to note that for large  $\tan\beta$
there is a sizeable minimum value for the R-parity 
breaking decay branching ratio.

\section{Discussion}

We have shown that in a class of supersymmetric models with 
R-parity breaking the stop can have new decay modes to third generation 
fermions such as $\stbt$. For stop masses accessible at LEP2 and 
smaller than the chargino masses the branching ratio for the
R-parity violating decay mode $\stbt$ can be comparable or even larger
than the R-parity conserving mode $\stch{1}$. 
Note also that quite generally the  R-parity breaking decay 
branching ratio could be increased through a larger value of 
$h_{\nu 33}$, a smaller value of $\delta$, a 
larger value of the Higgsino content of the neutralino, or 
any combination of these.
In fact, if R-parity is 
violated the lightest stop $\tilde{t}_1$ can be lighter than the lightest
neutralino $\chiz{1}$ leading to a $\sim  100~\%$ branching ratio into
the $\stbt$ decay mode.
Finally, one should not forget that even in the case when $\stch{1}$ 
is the main stop decay mode, the resulting signature will be different 
from that expected in the MSSM, because the lightest neutralino will 
decay. From \eq{vis}, (\ref{invi}), (\ref{invis}) the expected modes are:
$ \chiz{1} \to J + \nu,  3 \nu ,
l^- + l^+ + \nu, \tau^\pm + l^\mp + \nu,  
q + \bar{q} + \nu, q + \bar{q}' + \nu $.
While the first two decays give the same signature as in the MSSM,
which is a c-jet and missing momentum, the others give rise to
 events with  additional leptons or quarks.
In order to identify the corresponding signals in detail, one would have to
take into account all possible neutralino decay channels, 
and evaluate the corresponding detection efficiencies and standard
model backgrounds in a dedicated Monte Carlo simulation.

{\bf Acknowledgements}                      
                                    
This work was supported by DGICYT and {\sl Fonds zur F\"orderung der
wissenschaftlichen Forschung}, under grant numbers PB92-0084 and  
P10843-PHY, respectively. In addition, we have been supported
by Acci\'on Integrada Hispano-Austriaca (B\"uro f\"ur 
technisch-wissenschaftliche Zusammenarbeit des \"OAD'', 
project no. 1), as well as by Doctoral Fellowships from 
DGICYT (M. A. G. J.) and CNPq-Brazil (M. B. M.).

\noindent
\newpage



\bef
\centerline{
\psfig{file=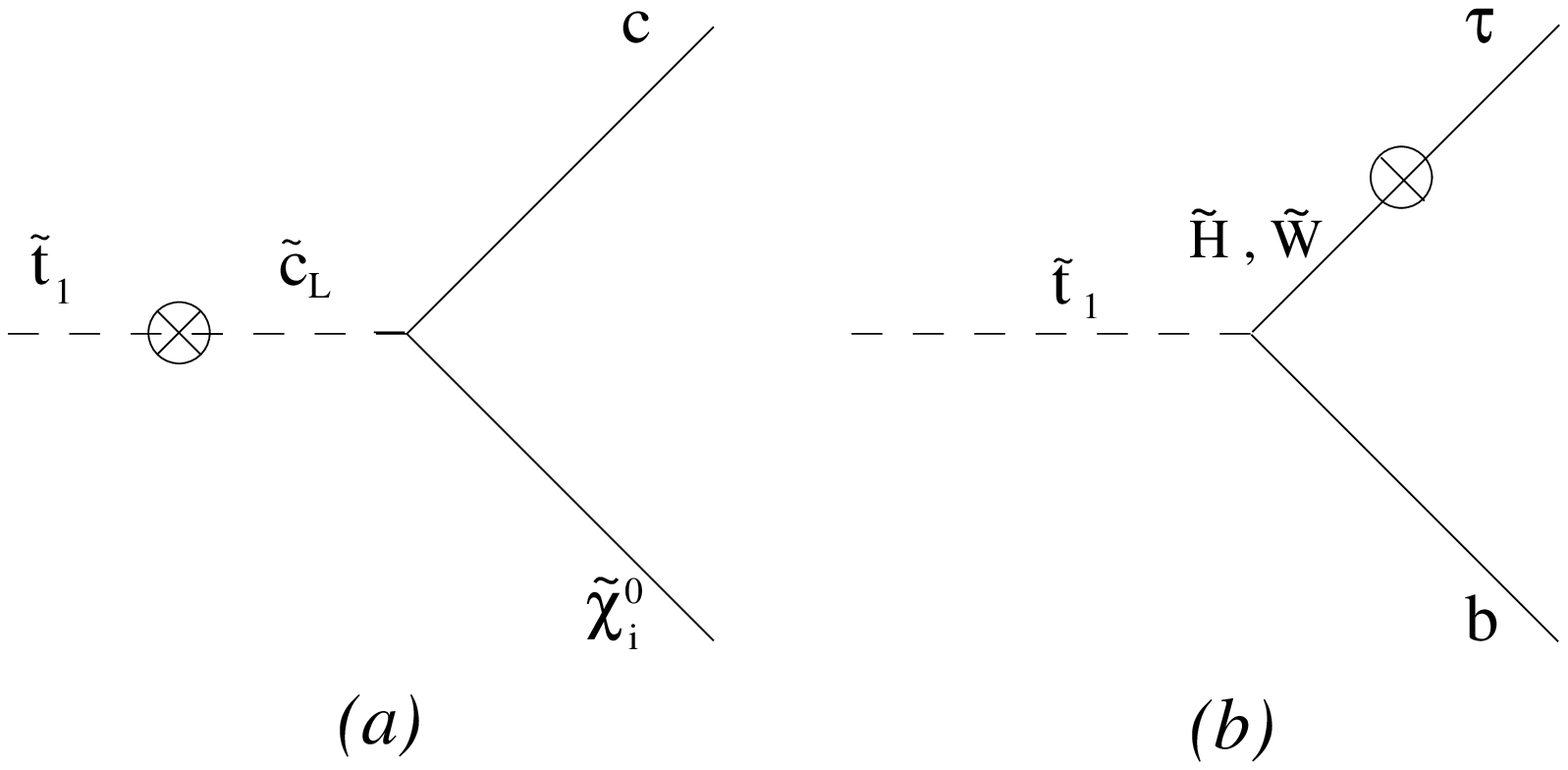,height=7cm,width=12cm}}
\caption{
Feynman diagrams leading to the standard R-parity conserving decay 
(a) and the new R-parity violating stop decay (b).
}
\label{stopdecay}
\eef


\bef
\centerline{
\psfig{file=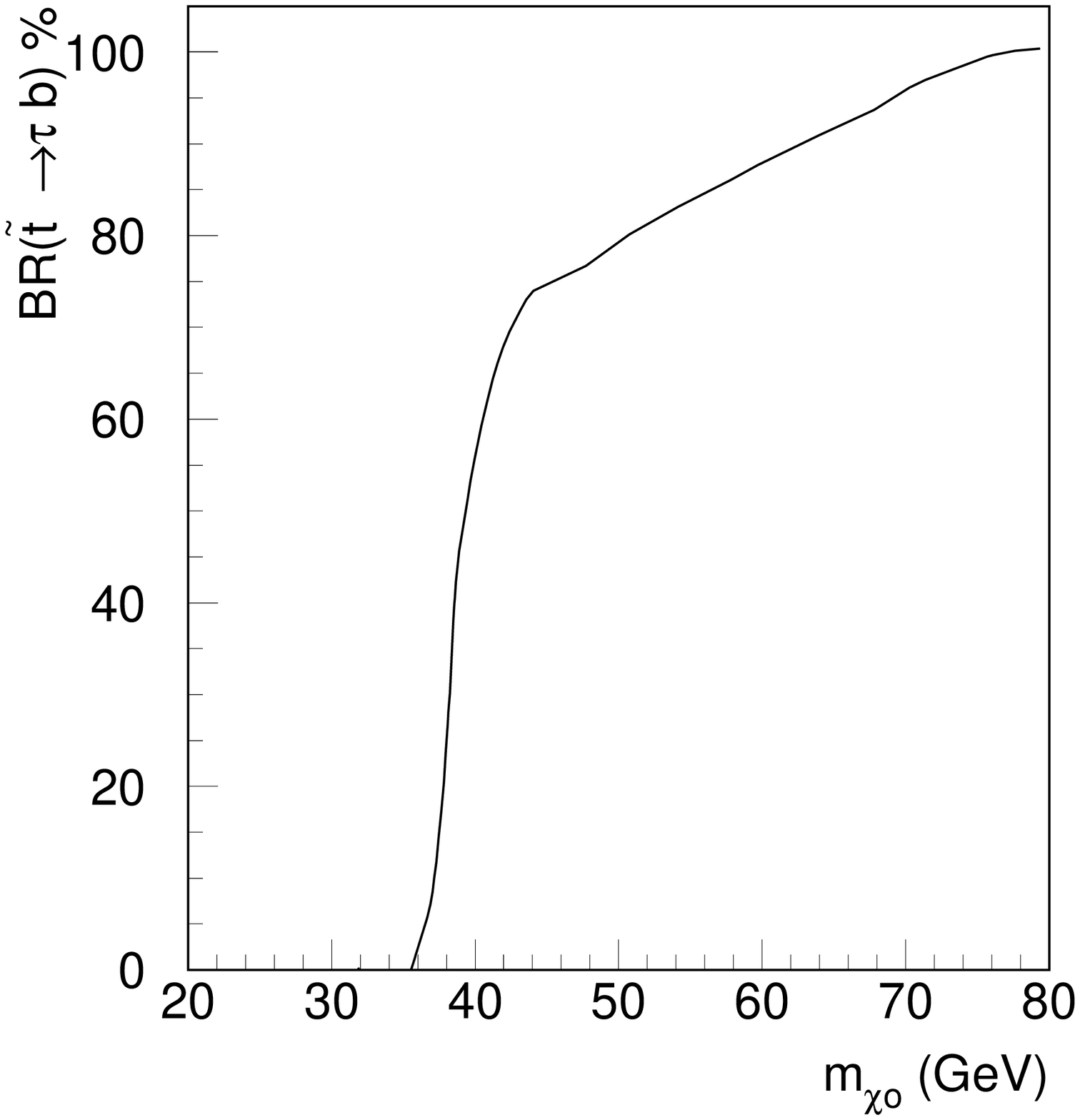,height=10cm}}
\caption{Barring unlikely small $f_1$ values as discussed
in the text, the curve displays the maximum values for the
branching ratio for $\stbt$ in per cent as a function of 
the lightest neutralino mass $\mchiz{1}$. We have taken
$\mst = 80$ GeV, $c_1 = 0.5$, $\delta = 0.1$, 
$v_R = 100$ GeV, $v_L = 0.1$ GeV and $h_{\nu 33} = 0.03$
and varied the remaining parameters as explained in the text}. 
\label{figbranch}
\eef

\bef
\centerline{
\psfig{file=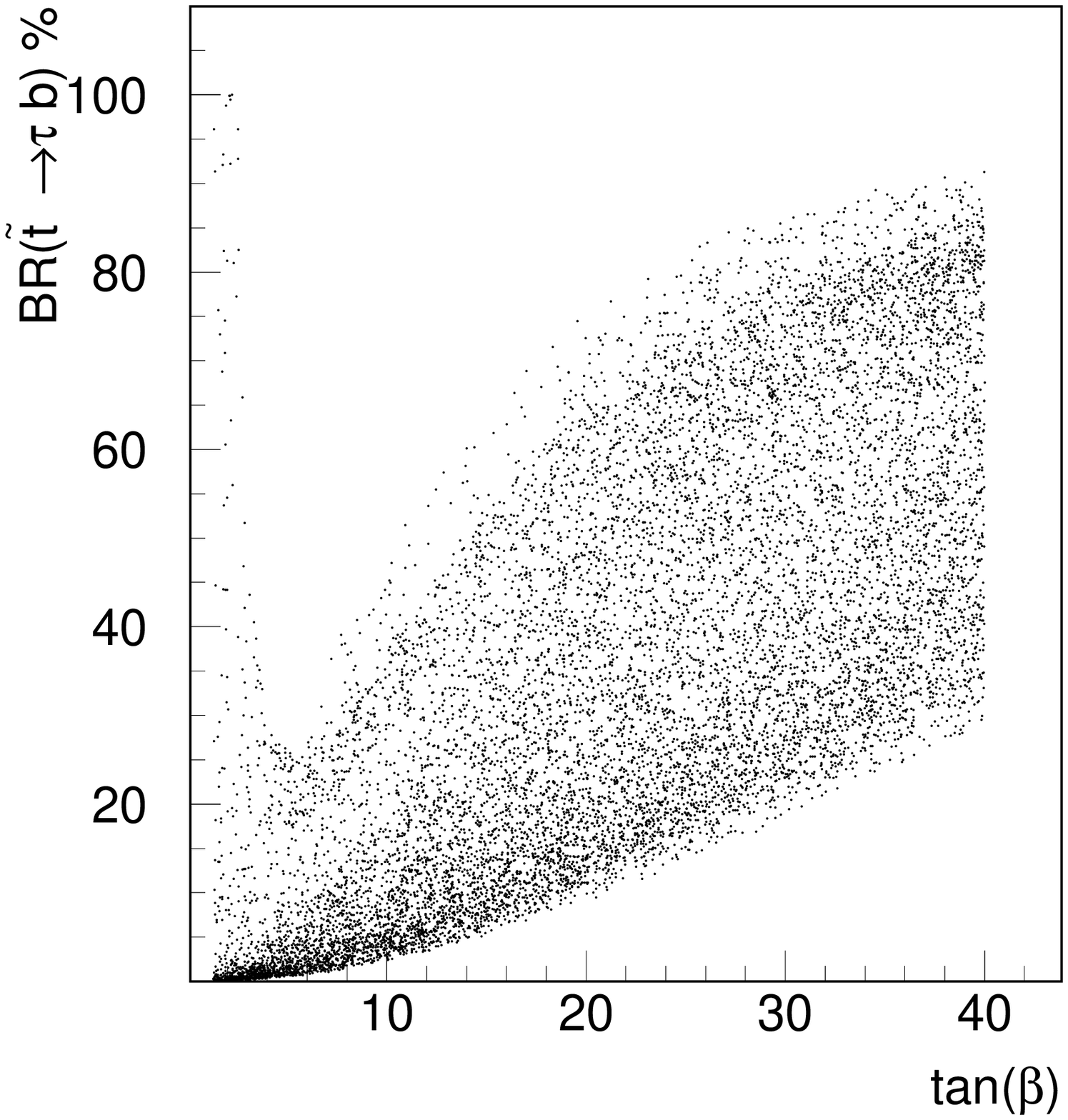,height=10cm}}
\caption{Scatter-plot for values of
branching ratio for $\stbt$ in per cent as a function of 
$\tan\beta$ for $55 < m_{\chi_1} < 60$ GeV. All parameters 
are chosen as in figure 2.}
\label{figtgbeta}
\eef

\end{document}